\documentclass[fleqn,usenatbib]{mnras}

\usepackage{newtxtext,newtxmath}

\usepackage[T1]{fontenc}

\DeclareRobustCommand{\VAN}[3]{#2}
\let\VANthebibliography\thebibliography
\def\thebibliography{\DeclareRobustCommand{\VAN}[3]{##3}\VANthebibliography}

\usepackage{graphicx}	
\usepackage{amsmath}	
\usepackage{multirow}
\usepackage{multicol}
\usepackage{array}

\newcolumntype{P}[1]{>{\centering\arraybackslash}p{#1}}
\newcolumntype{M}[1]{>{\centering\arraybackslash}m{#1}}

\title[Kroninan X-rays during rare planetary alignment]{Searching for Saturn's X-rays during a rare Jupiter Magnetotail Crossing using \textit{Chandra}}

\author[D. M. Weigt et al.]{
D. M. Weigt,$^{1}$\thanks{E-mail: D.M.Weigt@soton.ac.uk}
W. R. Dunn,$^{2}$
C. M. Jackman,$^{3}$
R. Kraft$^{4}$
G. Branduardi-Raymont,$^{2}$
J. D. Nichols,$^{5}$
\newauthor
A. D. Wibisono,$^{2}$
M. F. Vogt,$^{6}$
G. R. Gladstone$^{7,8}$
\\
$^{1}$School of Physics and Astronomy, University of Southampton, Southampton, UK\\
$^{2}$Mullard Space Science Laboratory, Department of Space and Climate Physics, University College London, Dorking, UK\\
$^{3}$School of Cosmic Physics, DIAS Dunsink Observatory, Dublin Institute for Advanced Studies, Dublin 15, Ireland\\
$^{4}$Harvard-Smithsonian Center for Astrophysics, Smithsonian Astrophysical Observatory, Cambridge, MA, USA\\
$^{5}$Department of Physics and Astronomy, University of Leicester, Leicester, UK\\
$^{6}$Center for Space Physics, Boston University, Boston, MA, USA\\
$^{7}$Space Science and Engineering Division, Southwest Research Institute, San Antonio, TX, USA\\
$^{8}$Department of Physics and Astronomy, University of Texas at San Antonio, San Antonio, TX, USA\\
}

\date{Accepted XXX. Received YYY; in original form ZZZ}

\pubyear{2021}

\begin{document}
\label{firstpage}
\pagerange{\pageref{firstpage}--\pageref{lastpage}}
\maketitle

\begin{abstract}
Every 19 years, Saturn passes through Jupiter's `flapping' magnetotail. Here, we report Chandra X-ray observations of Saturn planned to coincide with this rare planetary alignment and to analyse Saturn's magnetospheric response when transitioning to this unique parameter space. We analyse three Director's Discretionary Time (DDT) observations from the High Resolution Camera (HRC-I) on-board \textit{Chandra}, taken on November 19, 21 and 23 2020 with the aim to find auroral and/or disk emissions. We infer the conditions in the kronian system by looking at coincident soft X-ray solar flux data from the Geostationary Operational Environmental Satellite (GOES) and \textit{Hubble Space Telescope} (HST) observations of Saturn's ultraviolet (UV) auroral emissions. The large Saturn-Sun-Earth angle during this time would mean that most flares from the Earth-facing side of the Sun would not have impacted Saturn. We find no significant detection of Saturn's disk or auroral emissions in any of our observations. We calculate the 3$\sigma$ upper band energy flux of Saturn during this time to be  0.9~-~3.04~$\times$~10$^{-14}$~erg~cm$^{-2}$~s$^{-1}$ which agrees with fluxes found from previous modelled spectra of the disk emissions. We conclude by discussing the implications of this non-detection and how it is imperative that the next fleet of X-ray telescope (such as \textit{Athena} and the \textit{Lynx} mission concept) continue to observe Saturn with their improved spatial and spectral resolution and very enhanced sensitivity to help us finally solve the mysteries behind Saturn's apparently elusive X-ray aurora.

\end{abstract}




\section{Introduction}

The magnetospheres of Jupiter and Saturn are considered to be the two largest coherent structures in our Solar System with most of the plasma supplied by their moons and a variable interaction with the upstream solar wind \citep{Blanc2015,Bolton2015}. The magnetotail of Jupiter is so vast that it extends beyond Saturn's orbit at 9.5 AU (on average) \citep{Kurth1982, Lepping1982}, and contains a wide variety of plasma populations with different structures and velocities \citep{Mccomas2007}. \citet{Lepping1983} found that Jupiter's magnetotail ``flaps'' over a 2-3 day cadence and that the structure and movement of the tail are both determined by the variable solar wind dynamic pressure surrounding the jovian magnetosphere. Both \textit{Voyager} missions found that the tail had a density of $\sim$ 10$^{-3}$ - 10$^{-5}$ cm$^{-3}$ \citep{Gurnett1979a, Lepping1983, Kurth1982}, several orders of magnitude smaller than the denser, typical solar wind as observed close to Saturn ($\sim$0.01 - 0.5 cm$^{-3}$) \citep{Lepping1983}. As a result, the tail is theorised to resemble a ``sausage string'' carved by the solar wind with higher dynamic pressure creating the narrower structures of the tail and weaker pressure for the expanded regions. Every $\sim$ 19-20 years, the alignment of the gas giants is such that Saturn is located in Jupiter's wake with Saturn's magnetosphere alternately immersed in the solar wind and the rarefied deep jovian magnetotail. This phenomenon of ``overlapping'' planetary magnetospheric environments in the solar system is unique to Jupiter and Saturn due to the sheer scale of the jovian system. The Voyager 2 flyby of Saturn in 1981 occurred during the same planetary alignment which we observe in 2020. At that time, the evidence for the immersion of Saturn in Jupiter's tail came from the radio data, with almost complete dropouts of the Saturn Kilometric Radiation (SKR) observed \citep{Desch1983}. SKR emissions are observed at a frequency range of a few kHz to 1200~kHz on the dawn-noon sector of Saturn's auroral zone \citep{Lamy2009} and, like other planetary radio emissions (e.g. auroral kilometric radiation at Earth and jovian radio emissions), are believed to be produced from accelerated beams of electrons in the aurora via the cyclotron maser instability \citep{Zarka1998}. Further studies into SKR (e.g \citet{Reed2018}) found that these emissions can be used as a remote diagnostic tool to infer magnetospheric conditions at Saturn, as the timescale of many observed SKR events were found to correlate with possible auroral drivers (e.g. changing solar wind dynamic pressure or tail reconnection). \citet{Desch1983} observed during the flyby that the SKR emissions returned to a normal range of intensity when Saturn was back in the solar wind. The dropouts were therefore an indicator of intervals when Saturn was within Jupiter's magnetotail and were used as a tracer for the tail's apparent flapping motion. During this time, no observations of Saturn's auroral emissions in UV or X-ray wavelengths were taken and thus the global context for Saturn's magnetospheric response to such rare external conditions was not fully captured. 

The magnetospheric responses of both Jupiter and Saturn to various external (e.g. the solar wind) and internal (e.g. rapid rotation of the planet, moon-planet interactions) drivers \citep{Bagenal1992a} have been studied extensively using auroral measurements in the ultraviolet (UV) (e.g. \citet{Nichols2010,Nichols2016}), infrared (IR) (e.g. \cite{Badman2011}, \citet{Stallard2008, Stallard2008a}) and radio (e.g. \citet{Kurth2005}, \citet{Lamy2009}) wavebands. The processes are found to be highly complex in all wavebands and their auroral responses are varied (see detailed reviews by  \citet{Bhardwaj2000a} and \citet{Badman2015} for an overview of each of these processes). 

Compared to the other wavebands, the drivers of the X-ray auroral emissions remain the most mysterious. Auroral X-ray emissions are postulated to be present on Saturn since the discovery of its magnetosphere by the \textit{Voyager} spacecraft \citep{Opp1980, SANDEL1982} and the detection of its powerful auroral UV emissions at high latitudes \citep{Broadfoot1981}. The location and spectrum of these UV emissions suggested that they were caused by precipitating electrons with energies of $\sim$ 10s keV \citep{Carbary1983}. Such energetic emissions from electrons would suggest that X-ray bremsstrahlung processes would also be occurring. The first observation to try to detect the X-ray emissions from Saturn was carried out by the \textit{Einstein Observatory} on December 17 1979 \citep{Gilman1986}, prompted by the discovery of X-ray aurora on Jupiter \citep{Metzger1983}. From the $\sim$ 11ks observation of Saturn, no emissions were detected. \citet{Gilman1986} therefore concluded that bremsstrahlung was likely to be the dominant process for X-ray production. From this assumption, they calculated a 3$\sigma$ upper limit for the kronian X-ray flux at Earth to be 1.7~$\times$~10$^{-13}$~erg~cm$^{-2}$~s$^{-1}$. The value is compared with the expected energy flux detected at Earth, using a non-relativistic thick target bremsstrahlung model based on UV observations \citep{SANDEL1982}, of 8~$\times$~10$^{-16}$~erg~cm$^{-2}$~s$^{-1}$. This assumes that the thick-target bremsstrahlung occurs at high latitudes. The first significant detection of Saturn's X-rays was found by \textit{ROSAT} on April 30 1991 as part of a campaign to observe the gas and ice giants (Uranus and Neptune) \citep{Ness2000}. \citet{Ness2000} found that the upper limits of X-ray luminosites observed from Jupiter were much brighter that what was seen from the other 3 planets, which were found to have similar levels of weaker X-ray power. \textit{ROSAT} found the X-ray flux of Saturn to be greater than one order of magnitude than the Gilman et al's modelled bremsstrahlung flux. They concluded that this rise in flux may be due to a higher electron flux background that previous observations or an indicator of multiple X-ray mechanisms at Saturn in addition to X-ray bremsstrahlung. 

Subsequent X-ray campaigns with \textit{XMM-Newton} on 10 September 2002 \citep{Ness2004a} and \textit{Chandra} on 14 April 2003 \citep{Ness2004} discovered that, like Jupiter, Saturn has been observed to produce X-ray emissions from the planetary disk via elastic scattering of solar X-rays from its upper atmosphere.  Further \textit{Chandra} observations taken during January 2004 which coincided with an M6-class solar flare found that the disk X-ray flux increased by $\sim$ 5 times when the flare impacted Saturn \citep{Bhardwaj2005c}. The scattering mechanisms responsible for the disk emissions would require Saturn's disk to have a high X-ray albedo \citep{Ness2004}. \citet{Bhardwaj2005b} also found the X-ray spectrum of Saturn's rings was dominated by an atomic oxygen K$\upalpha$ fluorescence line at 0.53 keV, providing evidence for another X-ray mechanism on Saturn. \citet{Branduardi-Raymont2010} re-analyse the previous X-ray campaigns as well as two \textit{XMM-Newton} observations in 2005 and found that Saturn's disk emissions were well correlated with the solar cycle. The results from the \citet{Branduardi-Raymont2010} study verified what was found in previous literature and found the oxygen fluorescence line from the ring region varied differently from the disk emissions. During April-May 2011, \textit{Chandra} ACIS observations of Saturn were triggered to coincide with the predicted arrival of a solar wind shock, as determined from a propagated solar wind model \citep{Branduardi-Raymont2013}. In-situ \textit{Cassini} data during the \textit{Chandra} intervals confirmed the arrival of a solar wind shock at Saturn. \citet{Branduardi-Raymont2013} found X-ray emissions corresponding to scattering of solar X-rays in the upper atmosphere during an episode of flaring activity from the Sun. However, like all previous X-ray campaigns, there was no significant detection of X-ray aurora on Saturn. This led to the conclusion that more powerful solar wind shocks would be required to produce a signal above \textit{Chandra's} threshold of detectability. If shocks are acting on the planet, previous results would suggest that a mechanism producing X-ray bremsstrahlung may exist and are produced in a similar way as observed on Jupiter with photon energies $>$ 2 keV \citep{Branduardi-Raymont2008}, known as hard X-rays. These hard X-rays are emitted within the auroral region and have been observed to sometimes coincide with the main UV auroral oval. The brightest and most concentrated emissions are found from soft ($<$ 2 keV) X-ray photons which are often concentrated into a hot spot poleward of the main UV oval. The emissions have observed during many X-ray campaigns of Jupiter since their discovery by \citet{Gladstone2002} almost 20 years ago (e.g. \citet{Elsner2004, Branduardi-Raymont2007a, Branduardi-Raymont2007b}), and have been found to exhibit quasi-periodic pulsations (e.g. \citet{Dunn2016a, Wibisono2020, Weigt2020}). Such emissions are produced from charge exchange between precipitating ions, mainly from the volcanic moon Io, and the neutrals in the atmosphere (e.g. \citet{Cravens2003, Elsner2005, Ozak2010, Wibisono2020}). There has been no evidence so far of this type of mechanism occurring at Saturn although these emissions, as well as the theorised bremsstrahlung component, may be very dim and below the threshold of detection for both \textit{Chandra} and \textit{XMM-Newton} \citep{Hui2010a}.

In this paper, we report the first Chandra Director's Discretionary Time (DDT) observations of Saturn using Chandra's High Resolution Camera (HRC-I) as well as the first X-ray observations specifically designed to look at a planet's magnetospheric response during this rare planetary alignment. We assume that Saturn's magnetosphere will experience more powerful fluctuations as it moves from very rarefied plasma in Jupiter's tail to the denser solar wind (i.e. moving from densities of $\sim$ 10$^{-3}$ - 10$^{-5}$ cm$^{-3}$ to $\sim$ 0.01 - 0.5 cm$^{-3}$ of the typical solar wind). The \textit{Voyager 2} low density event plasma parameters also found during this time dynamic pressures of $\sim$ 10$^{-4}$ nPa \citep{Lepping1983}, unusually low values compared to the typical observed range of $\sim$ 0.01 - 0.1 nPa (e.g. \citet{Jackman2011}). These intervals of very low dynamic pressure will cause Saturn's magnetosphere to expand significantly and our study allows us to see what effect this dramatic change will have on the kronian magnetosphere. Therefore the shocks between the solar wind and rarefied tail will be stronger than what typically occurs at Saturn. We also look at data from GOES to monitor the solar activity during each of the observations. We conclude the paper by discussing our results and the implications of a non-detection of Saturn over the Chandra campaign and how the next generation of X-ray telescopes could help aid our understanding of Saturn's X-ray emissions.

\section{Observations and Data Analysis}

\subsection{Chandra}

\begin{table*}
    \caption{Chandra DDT observations of Saturn throughout November 2020}
    \label{table:hrc_obs}
    \begin{tabular}{ccccccc}
    \hline
    \multirow{2}{*}{\textbf{ObsID}} & \multicolumn{2}{c}{\textbf{Observation Start Date (UT)}} & \multirow{2}{*}{\textbf{Duration (ks)}} & \multirow{2}{*}{\textbf{\begin{tabular}[c]{@{}c@{}}Apparent\\ diameter\end{tabular}}} & \multirow{2}{*}{\textbf{\begin{tabular}[c]{@{}c@{}}Heliocentric\\ distance (AU)\end{tabular}}} & \multirow{2}{*}{\textbf{\begin{tabular}[c]{@{}c@{}}Chandra-Saturn\\ distance (AU)\end{tabular}}} \\
     & \textbf{dd/mm/yyyy} & \textbf{(HH:MM)} &  &  &  &  \\ \hline
    \textbf{24845} & 19/11/2020 & (11:27) & 10.0 & 15.90\arcsec & 9.99 & 10.45 \\
    \textbf{24846} & 21/11/2020 & (12:09) & 10.01 & 15.85\arcsec & 9.99 & 10.48 \\
    \textbf{24847} & 23/11/2020 & (22:44) & 9.89 & 15.80\arcsec & 9.99 & 10.52 \\
    \hline
    \end{tabular}
\end{table*}

We report the analysis of three $\sim$ 10 ks \textit{Chandra} HRC-I X-ray DDT observations of Saturn (ObsID 24845, 24846 and 24847) taken place during mid-November 2020. The start date, duration, apparent diameter of Saturn, heliocentric and Chandra-Saturn distance are shown for each observation in Table~\ref{table:hrc_obs}. The observations were taken in $\sim$ 2-day cadence to account for the motion of Jupiter's magnetotail, immersing Saturn every 2-3 days \citep{Lepping1983}. All three observations were acquired using Chandra HRC-I with the movement of Saturn across the chip corrected for in each observation prior to analysis using the Python equivalent of CIAO's \verb'sso_freeze' algorithm \citep{Weigt2020}.

From previous \textit{Chandra} ACIS and \textit{XMM-Newton} observations of Saturn, disk X-ray emissions have been detected with high statistical significance. The spectrum of the disk emission is well fitted by an optically thin coronal model with average temperature of $\sim$ 0.5 keV, resulting from the scattering of solar X-rays in Saturn's upper atmosphere \citep{Branduardi-Raymont2010}. When the fluorescent oxygen emission line found by \citet{Bhardwaj2005c} at 0.53 keV was included in the fit, it was found to significantly improve it. At the time of writing, there have been no statistically significant detections of auroral X-ray photons at Saturn.

The contaminant build-up on ACIS now severly inhibits the detection of X-rays below 1 keV (see \citet{Plucinsky2018}), where we expect the peak X-ray photon energy from Saturn to lie. Therefore these observations were conducted with the High Resolution Camera (HRC). The HRC-I detector on-board \textit{Chandra} has no spectral resolution, resulting in the X-ray photons we analyse being unfiltered into energy bins. This therefore makes it difficult to differentiate between the different possible processes responsible for X-ray production (e.g. bremsstrahlung, solar X-ray scattering) at Saturn. The main focus of these observations however, was to utilise the high spatial resolution of HRC-I to allow us to separate clearly auroral and disk X-ray regions at Saturn, similar to what can be done at Jupiter. The spatial resolution of HRC-I is $\sim$ 50$\%$ better than that of ACIS allowing us to map any detectable X-ray emissions onto Saturn's surface in greater detail. Once mapped, the intensity of the X-rays are calculated from the 0.8\arcsec full-width at half-maximum (FHWM) of the HRC-I point spread function (PSF) as well as the assumptions we consider about the altitude of the emissions in Saturn's ionosphere (more details of the mapping algorithm can be found in \citet{Weigt2020}).

\subsection{GOES}
To help establish a possible correlation between solar activity and X-ray flux during this time, as found by \citet{Bhardwaj2005c} and \citet{Branduardi-Raymont2010}, we use data from the Geostationary Operational Environmental Satellite (GOES) \citep{Lemen2004}. GOES is used to monitor the effect weather systems have on Earth as well as space weather effects from the Sun, using instruments such as the GOES X-ray Sensor (XRS). We used data from XRS on GOES-16 \citep{Goodman2013} to monitor any flaring activity from the Sun. The XRS monitors both soft (1 \r{A} - 8 \r{A}) and hard (0.5 \r{A} - 4 \r{A}) solar X-rays and measures the flux emitted from the Sun. The magnitude of the soft X-ray flux is then used to identify the type of solar flare observed from an A-class (weakest flare: 10$^{-8}$ W m$^{-2}$) to an X-class flare (strongest flare: $\geq$ 10$^{-4}$ W m$^{-2}$). 

When using the GOES data, we take into account the Sun-Earth and Sun-Saturn-Earth light time to ensure we have the correct time for when the radiation from the flare was detected by GOES and impacted Saturn respectively (i.e. as viewed by \textit{Chandra} and \textit{HST}). This was carried out for each of the Chandra DDT observations. We also consider the Saturn-Sun-Earth angle during each of the Chandra observations. This allows us to get an idea of the direction of any flares found by GOES. If this angle is large, the uncertainty of a solar flare found from any detectors near the Earth-facing side of the Sun impacting Saturn will also be large.

\section{Results}

\subsection{Chandra}

\begin{figure}
	\includegraphics[width=\columnwidth]{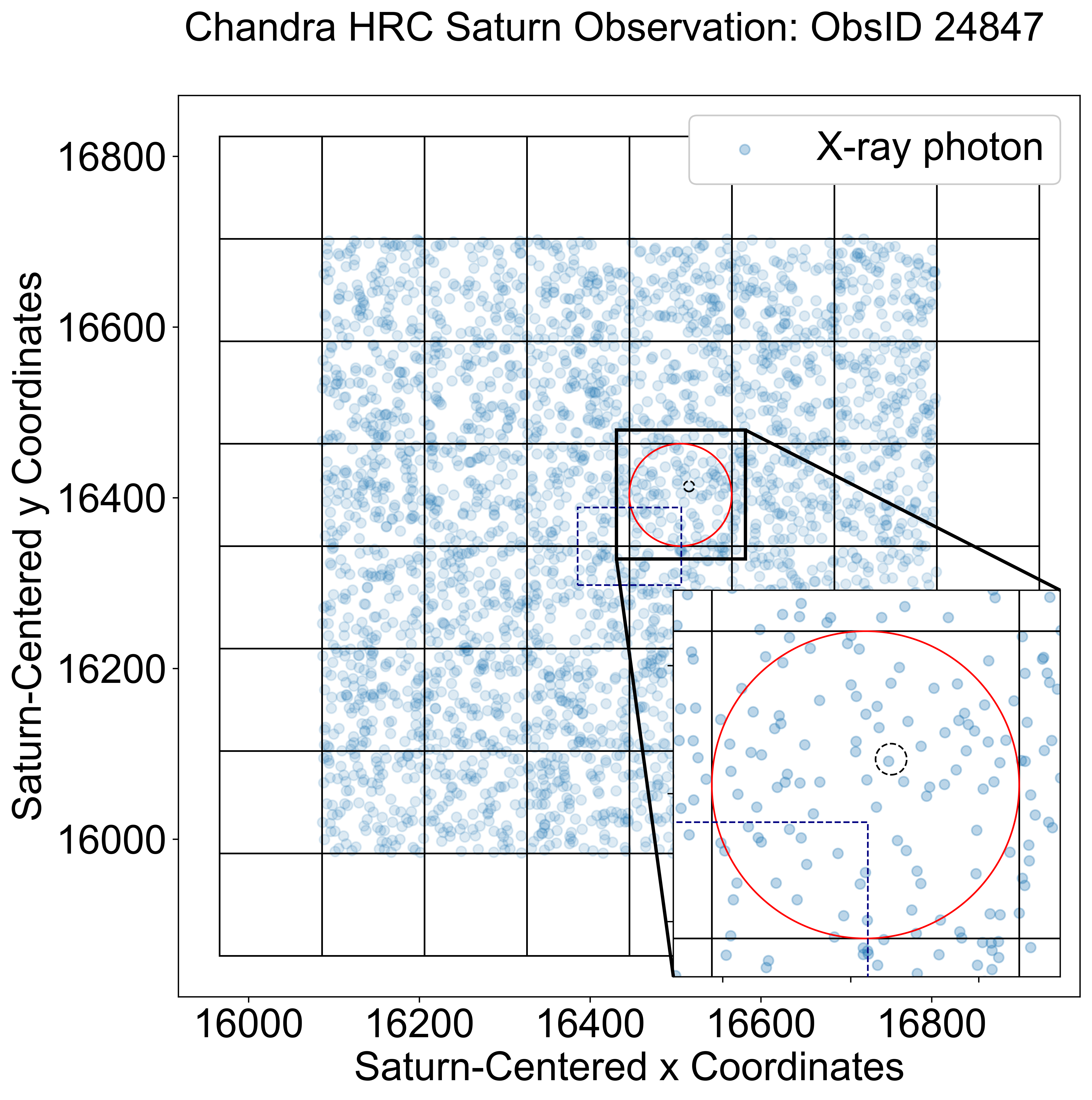}
    \caption{Positions of all X-ray photons (blue dots) on the HRC-I~chip, observed during ObsID 24847. The aim-point of the HRC detector is located at the centre of the grid, surrounded by a navy-dashed box (16\arcsec $\times$ 12\arcsec) indicating the error associated with the effective aim-point centre. The PSF of the detector is shown by the black-dashed circle. The chip is split into a grid with each box the same width as Saturn's disk during the observation. The corrected position of Saturn on the chip is found to be near the center, as shown by the red circle. The number of photons in each box are then compared with those found within Saturn to find the significance of the X-ray detection on the disk. Bottom right corner shows the Saturn region in more detail. The position of all photons on the grid have been converted to planetocentric (i.e. Saturn centered) coordinates on the HRC-I~ chip. The same grid was produced for all 3 observations.}
    \label{fig:hrc_grid}
\end{figure}

We expect a low-count regime at Saturn based on previous observations. Following \citet{Dunn2021}'s detection of X-rays from Uranus, we divide the detector into a grid of cells the size of the planet. The algorithm first assesses the significance of the detection by splitting the chip into a grid with cells the size of the region of interest i.e. Saturn's disk. Fig.~\ref{fig:hrc_grid} shows the position of Saturn on the chip using the apparent angular diameter of $\sim$ 16\arcsec, RA and DEC from Saturn ephemeris data (JPL Horizons) to create a region using SAOImage DS9 \citep{Joye2003} for ObsID 24847. The region is then over-plotted onto the image as shown by the red circle in Fig.~\ref{fig:hrc_grid}. The counts within the Saturn region are then compared to the background counts on the remainder of the chip. The X-ray photons shown in Fig.~\ref{fig:hrc_grid} are for ObsID 24847 (23 November 2020) as this observation had the greatest number of counts of the three Saturn observations. The aim-point of the detector is located at the chip centre, surrounded by an error box (navy-dashed 16\arcsec $\times$ 12\arcsec box) accounting for possible drifts of the effective aim-point centre. As shown in Fig.~\ref{fig:hrc_grid}, the error in the pointing will have a minimal effect on our observations. The PSF of the detector is represented by a black-dashed circle to show our resolution when observing Saturn's disk.  
 
\begin{figure*}
	\includegraphics[width=\textwidth,height=7cm]{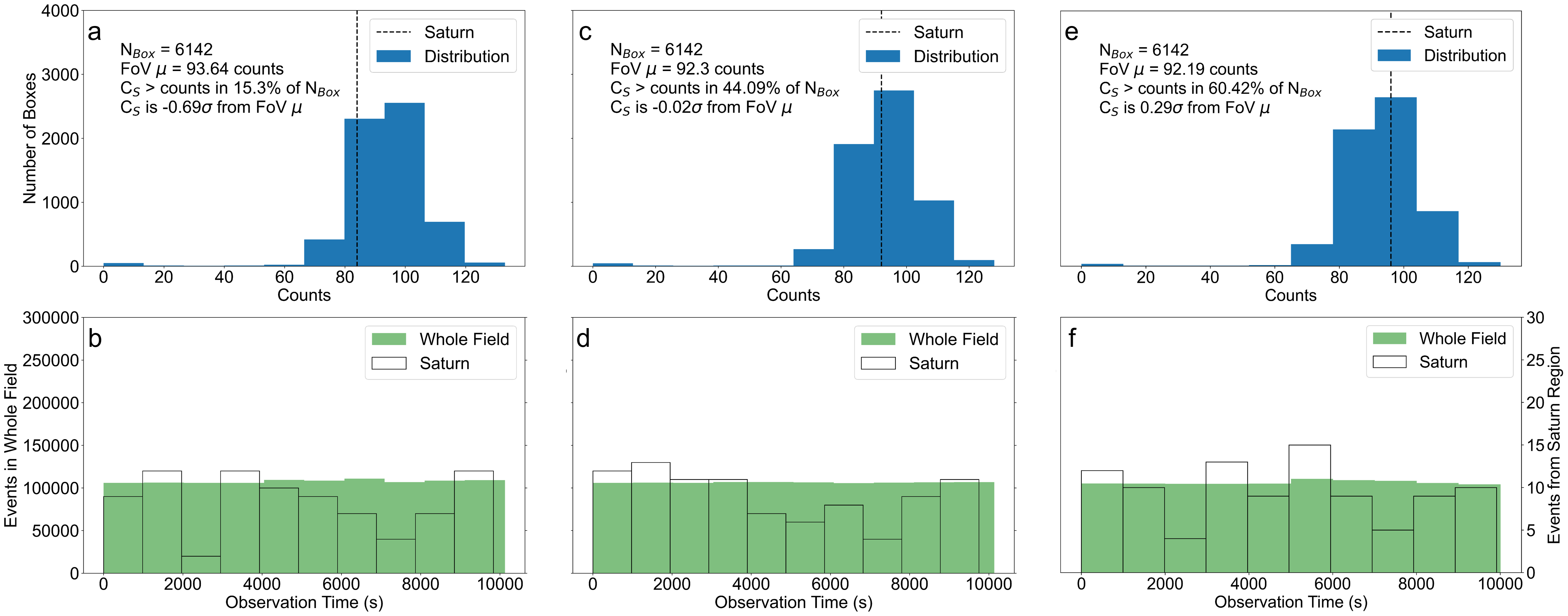}
    \caption{The histogram distribution of counts from each Saturn-disk sized box on the grid (Fig.~\ref{fig:hrc_grid}) relative to the number of counts from the Saturn region for (a) ObsID 24845, (c) 24846 and (e) 24847. The corresponding time distributions within the Saturn region (transparent distribution, with $y$-axis on the right) compared to the the whole distribution on the chip (green) for all observations are shown in panels (b), (d) and (f) respectively. The counts found from within the Saturn region, $C_{S}$, are represented by the black-dashed line in the top panel overlaid on the full count distribution. Shown in text within panel a) are the total Saturn-disk sized boxes across the detector ($N_{Box}$), the Field of View mean found within the boxes ($FoV$~$\mu$) or background counts on the chip, the percentage of $N_{Box}$ that contained a number of counts smaller than $C_{S}$ and the number of standard deviations ($\sigma$) $C_{S}$ was from the $FoV$~$\mu$.}
    \label{fig:p_and_t}
\end{figure*}

We do not detect Saturn in any of the November 2020 observations. Each of the observations shows no significant detection ($<$ 3$\sigma$ detection from background, where $\sigma$ is the standard deviation) of auroral or disk X-rays from Saturn during this unique event. The distributions of all counts across the chip within each Saturn-disk sized box (Fig.~\ref{fig:hrc_grid}) relative to the counts within the Saturn region for each observation are shown in Fig.~\ref{fig:p_and_t}a), (c) and (e). Each observation had a similar distribution, with the number of Saturn counts located at the peak of the distribution. This therefore shows that the signal-to-noise ratio was too low for each of the DDT observations and we cannot distinguish Saturn from the background or $FoV$~$\mu$. As shown in Table~\ref{table:cxo_pow_wow} and the histogram distribution of counts in Fig.~\ref{fig:p_and_t}a), (c) and (e), the mean background or Field of View counts on the HRC-I~chip ($FoV$~$\mu$) were comparable to the observed counts found within the Saturn region ($C_{S}$: black-dashed line). The total number of boxes across the chip (denoted as $N_{Box}$) used in the distribution, the $FoV$~$\mu$, the percentage of $N_{Box}$ that had fewer counts than $C_{S}$ and the number of standard deviations $C_{S}$ is from the $FoV$~$\mu$ are all displayed in Fig.~\ref{fig:p_and_t}a), (c) and (e). The total number of Saturn counts and the resulting net counts, $C_{S} - FoV$~$\mu$ for all three observations are displayed in Table~\ref{table:cxo_pow_wow}. Assuming Poissonian statistics and accounting that the kronian X-rays photons are not distinguishable from the background, we calculate the error on the net counts found from  ObsID 24847, the observation with the highest $C_{S}$, to be 13.71 cts. These assumptions were used for the remaining \textit{Chandra} observations (as shown in Table~\ref{table:cxo_pow_wow}).

We also investigate the time series across the HRC-I~chip to search for any time variability in the X-rays registered from the region commensurate with Saturn's position compared to the full distribution. Analysis of similar light curves for Jupiter's X-rays shows features including: disk X-rays which broadly correspond to the solar X-ray activity levels (e.g. \citet{Bhardwaj2005}, \citet{Dunn2020a}), auroral X-rays which show distinct hot spots in the north and south (e.g. \citet{Gladstone2002}, \citet{Dunn2017}) as well as quasi-periodic flaring on rare occasions \citep{Jackman2018a, Weigt2021}. For the Saturn case (Fig.~\ref{fig:p_and_t}b), (d) and (f)), the light curve for the entire chip is almost completely flat for all observations, and the subset of this light curve which emerges from the Saturn location is not significant above background.   

\begin{table*}
\caption{Chandra DDT observation results from the low count algorithm with upper limit power estimates}
\label{table:cxo_pow_wow}
    \begin{tabular}{ccccccccl}
    \cline{1-8}
    \textbf{ObsID} & \textbf{\begin{tabular}[c]{@{}c@{}}Counts in Saturn\\ region ($C_{S}$)\end{tabular}} & \textbf{\begin{tabular}[c]{@{}c@{}}Mean Background\\ Counts \\ ($FoV$~$\mu$)\end{tabular}} & \textbf{\begin{tabular}[c]{@{}c@{}}Net Counts\\ $\left(\pm \sqrt{(C_{S} + FoV~\mu)}\right)$\end{tabular}} & \textbf{\begin{tabular}[c]{@{}c@{}}3$\sigma$ upper limit$^{a}$ \\ Disk power (GW)\end{tabular}} & \textbf{\begin{tabular}[c]{@{}c@{}}3$\sigma$ upper limit$^{a}$\\ Disk Flux$^{b}$ \end{tabular}} & 
    \textbf{\begin{tabular}[c]{@{}c@{}}Mean GOES \\ soft X-ray flux\\ ($\times$ 10$^{-7}$ Wm$^{-2}$) \end{tabular}} & \\ \cline{1-8}
    \textbf{24845} & 84 & 94 & -10 $\pm$ 13.34 & 0.28 & 0.90 & 0.83 & \\
    \textbf{24846} & 92 & 92 & 0 $\pm$ 13.56 & 0.76 & 2.48 & 3.17 & \\
    \textbf{24847} & 96 & 92 & 4 $\pm$ 13.71  & 0.95 & 3.04 & 6.93 & \\
    \cline{1-8}
    \multicolumn{8}{l}{$^{a}$ calculated 3$\sigma$ upper limit as all observations were non-detections.} \\
    \multicolumn{8}{l}{$^{b}$ upper limit of X-ray disk flux has units $\times$10$^{-14}$ erg cm$^{-2}$  s$^{-1}$} \\
    \end{tabular}
\end{table*}

From the net counts in Table~\ref{table:cxo_pow_wow}, we calculate the 3$\sigma$ upper limit of the disk emissions during this time. As this is the first auroral observation campaign linked to the alignment of Jupiter and Saturn, this is the first opportunity to predict X-ray powers for this very special case and test them against measurements from previous studies. The true mean for the upper limit is found from the average value of the parameter from the range of net counts for each observation. The energy flux calculation used here assumes a photon energy of $\sim$ 0.5 keV, the average temperature of the X-ray emissions found by \citet{Branduardi-Raymont2010}, and is an estimate of the flux detected from Saturn. This approach has been used for previous HRC-I observations of Jupiter \citep{Gladstone2002,Dunn2017,Weigt2020,Weigt2021}. The upper limit of the emitted disk power was found by multiplying the fluxes by 4$\pi~d^{2}$, where d is the average geocentric distance of Saturn during each observation. The 3$\sigma$ upper limit power of the disk from the observation with the highest number of Saturn counts, ObsID 24847, is 0.95 GW, corresponding to an energy flux upper limit of 3.04~$\times$~10$^{-14}$~erg~cm$^{-2}$~s$^{-1}$, assuming the net counts corresponded to Saturn X-ray emissions. The larger errors in net counts are a result of the high background throughout each observation, mainly from high energy galactic cosmic ray photons, which are found to exceed or equal the source counts. The upper limits of calculated for all observations are also shown in Table~\ref{table:cxo_pow_wow}.

\subsection{Solar data}
In the absence of an in situ monitor (i.e. a spacecraft in orbit at or near Saturn) we must rely on other remote measurements to infer the conditions in the kronian system at the time of the Chandra observations. One way of doing this is through solar observations as we expect, by analogy with Jupiter \citep{Bhardwaj2005} and previous X-ray studies of Saturn \citep{Bhardwaj2005c, Branduardi-Raymont2010}, that the disk power may track the solar X-ray flux as found by GOES. The solar soft X-ray flux is found to increase, as well as other wavebands across the electromagnetic spectrum, when the Sun emits a solar flare \citep{Fletcher2011}. In this study, we use the GOES solar soft X-ray flux as a diagnostic for solar X-ray activity. The mean GOES solar X-ray flux during each observation is shown in Table~\ref{table:cxo_pow_wow}.

To indirectly infer the state of the kronian magnetosphere, one could use auroral observations in other wavelengths. A \textit{Hubble Space Telescope} (\textit{HST}) UV campaign (P.I J. Nichols) has selected observations from October 29 until November 23 2020, to try and observe the UV auroral response to Saturn being in and out of Jupiter's tail. From this HST campaign, there were 2 \textit{HST} observations within 2 days of the Chandra observations on November 22 $\sim$ 08:20 UT and November 23 $\sim$ 05:00 UT, $\sim$ 18 hours before ObsID 24847. These observations may indicate that the dominant process producing the UV aurora was different between both observations with a dramatic change of morphology from a spiral-like structure to more localised brightening near the pole surrounded by half the UV oval (discussed in more detail in Nichols et al., (in prep.)). This change in auroral morphology may suggest that the magnetosphere was changing state over short timescales.

\begin{figure}
	\includegraphics[width=\columnwidth]{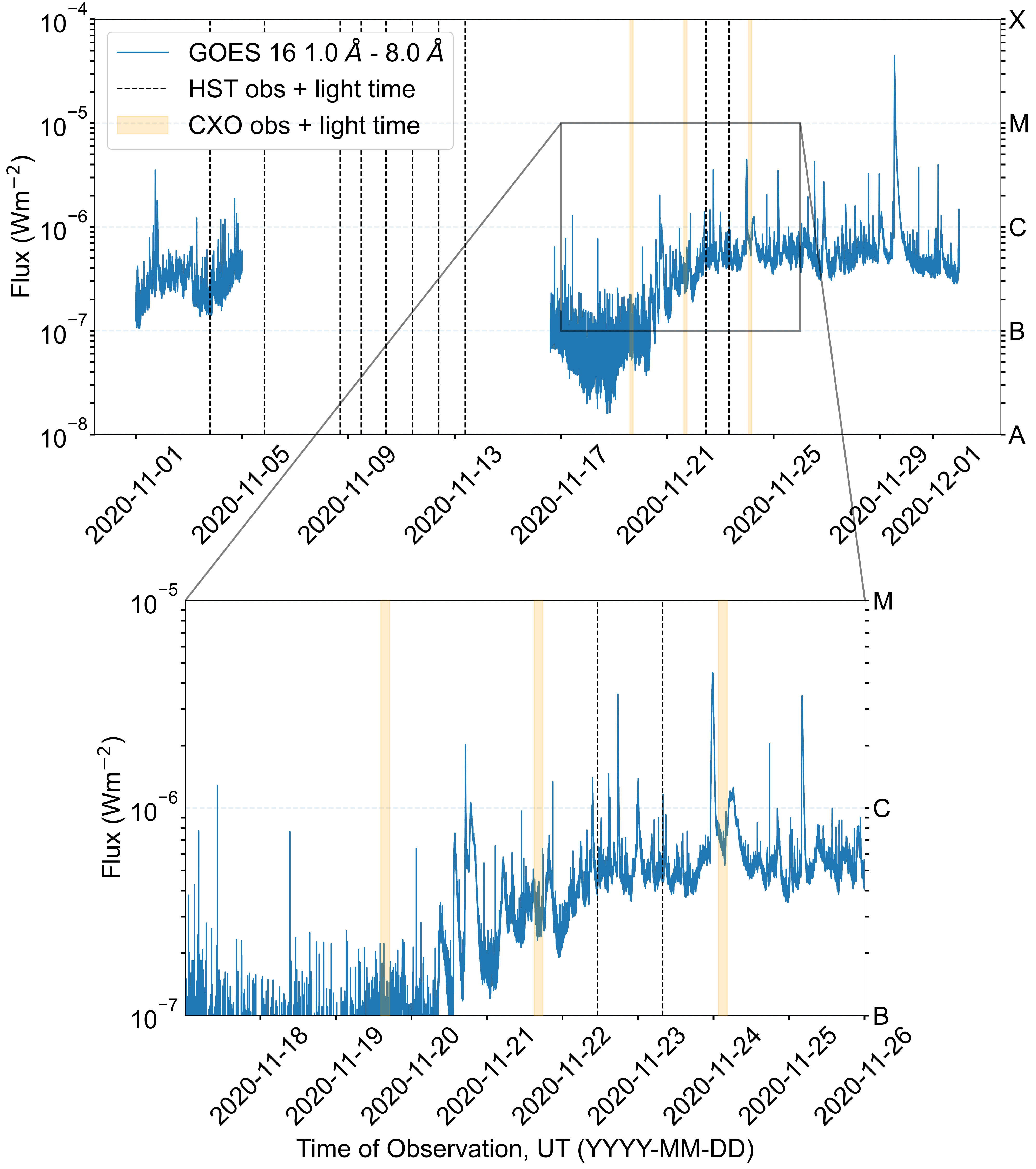}
    \caption{Plots of the observation time vs soft solar X-ray flux in logspace from GOES 16 for (a) the full month of November and (b) the week encompassing the Chandra observations. There was no GOES data recorded between November 5 and November 16. Both \textit{HST} (black dashed line) and \textit{Chandra} (shaded orange regions) observation intervals are plotted in each panel. The Sun-Earth and Sun-Saturn-Earth light times were used to account for the flare being observed by GOES and when it impacted Saturn and viewed by \textit{Chandra} and \textit{HST} respectively. The ObsID 24847 observation window is shown in both panels by the final shaded area. The estimated arrival time of the radiation from the C-class flare is found to impact Saturn just prior to or coincide with ObsID 24847 as shown in both panels. The C-class flare is found to have a flux of $\sim$ 4 $\times$ 10$^{-6}$ W m$^{-2}$. This is the peak solar flux observed during throughout the Chandra campaign after a steady increase of average solar flux throughout November (panel (a)).}
    \label{fig:goes}
\end{figure}

The soft X-ray flux from GOES and the observation intervals of both \textit{HST} (dashed black lines) and the three \textit{Chandra} observations (shaded orange regions) are shown in Fig.~\ref{fig:goes} for the entire month of November (panel a)) and the week encompassing the \textit{Chandra} observations (panel (b)). We note that there was no GOES 16 data from November 5 until November 16. As shown in Fig.~\ref{fig:goes}, the GOES solar soft X-ray flux was found to increase from around November 20 and plateau at an overall higher average solar flux from November 23. This is also shown in Table~\ref{table:cxo_pow_wow} with the increase in mean solar X-ray flux throughout the \textit{Chandra} campaign. This flux level is then observed to be maintained throughout the remainder of the month (Fig.~\ref{fig:goes}b)). The solar flux during our observations indicates a possible C-class flare coinciding with or arriving at Saturn just before ObsID 24847. The Sun-Earth and Sun-Saturn-Earth light times were used to correct for the GOES detection of the flare (light time ($lt$) $\sim$ 8.2 min) and for when the flare would impact Saturn as viewed by \textit{Chandra} and \textit{HST} ($lt$ $\sim$ 170 min) and used in all intervals shown in both panels. However, the Saturn-Sun-Earth angles throughout the campaign were inferred to be $\sim$ 72$^\circ$ - 89$^\circ$, suggesting that any flares found from GOES would unlikely to have impacted Saturn. The flux of the flare is $\sim$ 4 $\times$ 10$^{-6}$ W m$^{-2}$ and, although unlikely to have arrived at Saturn, is found to raise the mean solar flux for the remainder of the campaign. As observed on Jupiter and from previous studies of Saturn, not all of the incident solar photons will be scattered back due to the planet's X-ray albedo \citep{Bhardwaj2005, Ness2004a, Ness2004}. The X-ray albedo of Saturn is predicted to be $>$ 5.7 $\times$ 10$^{-4}$ (slightly higher than Jupiter's at $\sim$ 5 $\times$ 10$^{-4}$) which means that Saturn's upper atmosphere acts like a slightly opaque, diffuse mirror for the incoming solar X-rays, scattering back $\sim$ 1 in a few thousand photons. This would allow Saturn to be observed in the soft X-ray energy range, as found previously, and one would expect the increase in solar flux should allow a significant detection of these emissions.

\section{Discussion and Conclusions}

As we have discussed, the first \textit{Chandra} HRC-I DDT observations of X-rays from Saturn during this unique alignment with Jupiter show a non-detection of Saturn (both disk and auroral emissions). We note that the observations occurred during the early rising phase of the solar cycle, when the Sun displays generally low solar X-ray flux, sporadic flaring and is ejecting short scale fast solar wind streams (e.g. \citet{Atac2001, Burlaga2001, Xystouris2014}). The solar flux during the DDT campaign only peaked above 10$^{-6}$ Wm$^{-2}$ during one of the three observation windows (ObsID 24847) (Fig.~\ref{fig:goes}), immediately after a C-class flare. Due to low solar X-ray fluxes and the large Saturn-Sun-Earth angle throughout we would expect the disk emissions to be very dim as there will be fewer solar photons undergoing elastic scattering in Saturn's upper atmosphere. This would therefore explain the non-detection of Saturn's disk amongst the background (dominated by high energy galactic cosmic rays) of the chip. We do note however that the observation found to have the most Saturn counts (C$_{S}$), ObsID 24847, is found to have the highest observed mean GOES flux. This agrees with previous studies that have found a similar correlation between the number of disk X-ray photons and solar X-ray flux (e.g. \citet{Bhardwaj2005c}). 

The calculated 3$\sigma$ range of upper limit disk energy flux in this study (0.90 - 3.04~$\times$~10$^{-14}$~erg~cm$^{-2}$~s$^{-1}$), using our assumptions, is found to be within the upper end of the range of energy fluxes from modelled spectra of the \textit{Chandra} disk emissions between 0.2 and 2 keV found by \citet{Branduardi-Raymont2010} ($\sim$ 0.30 - 1.00~$\times$~10$^{-14}$~erg~cm$^{-2}$~s$^{-1}$). These observations were found to coincide with GOES soft X-ray flux measurements exceeding 10$^{-6}$ Wm$^{-2}$, similar to the solar activity measured during November 2020. Our energy flux upper limits are an order of magnitude less than the 3$\sigma$ upper limit flux calculated by \citet{Gilman1986} for high latitude, non-relativistic thick-target bremsstrahlung at 1.7~$\times$~10$^{-13}$~erg~cm$^{-2}$~s$^{-1}$, from their $\sim$ 11ks observation. When compared to the results of \cite{Ness2000}, our 3$\sigma$ upper limit flux range lies within both their observed Saturn soft X-ray flux (1.9~$\times$~10$^{-14}$~erg~cm$^{-2}$~s$^{-1}$) and 95$\%$ confidence upper limit flux for hard X-ray flux (1.3~$\times$~10$^{-14}$~erg~cm$^{-2}$~s$^{-1}$). Therefore our predicted fluxes are consistent with previous observations and predictions. We note that unlike previous \textit{Chandra} observations, no spectra were taken and therefore no energy filtering was carried out for the DDT observations. The estimated power calculated here will therefore include photon energies exceeding the 2 keV upper energy bound of the models used by \citet{Branduardi-Raymont2010}. We therefore suggest that future campaigns look again at Saturn during this rare planetary alignment with the next fleet of X-ray telescopes with greatly improved X-ray sensitivity such as \textit{Athena} \citep{Barret2016} and \textit{Lynx} \citep{Gaskin2019}, one of the mission concepts currently under review for the 2020 Astrophysics Decadal Survey. The collecting area at 1 keV of both \textit{Athena} ($\sim$ 1.5 m$^{2}$) and \textit{Lynx} ($\sim$ 2.3 m$^{2}$) is $\sim$ 20 - 30 times greater than Chandra ($\sim$ 0.08 m$^{2}$) which allows for deeper exploration of space by increase the limit of detection of fainter sources (assuming a relatively low background). The combination of the large collecting area of \textit{Lynx} with a high spatial resolution ($<$ 0.5\arcsec) will allow us to map the fainter individual X-ray photons in greater detail. Although the spatial resolution of \textit{Athena} ($\sim$ 5\arcsec) is insufficient to allow detailed mapping of the X-ray emissions, the spectral resolution of \textit{Athena's} \textit{X-ray Integral Field Unit (X-IFU)} is $\sim$ 2.5 eV up to energies of 7 keV \citep{Barret2016}, surpassing both the \textit{High Definition X-ray Imager} on board \textit{Lynx} ($\sim$ 70 eV - 150 eV at energies 0.3 - 5.9 keV) and ACIS ($\sim$ 130 eV - 280 eV at energies 1.49 - 5.9 keV) \citep{Falcone2019}. The high energy resolution of \textit{Athena} will allow us to find more detailed spectra of the X-ray emissions at Saturn as well as that of the other planets. With the greater sensitivity and spectral resolution of the new X-ray fleet, we would have the capabilities of conducting a deeper search for X-ray aurorae on Saturn. This will allow us to further constrain the disk power and to provide a more accurate estimate of Saturn's disk response to such conditions as well as other planets, further aiding our understanding of the X-ray production mechanisms within our Solar System. However, to ensure we get a significant number of kronian X-ray photons (and planets producing fainter X-ray emissions) to utilise the spectral and spatial capabilities of these telescopes, future observations will need to optimise the exposure time required. 

As stated previously, this complicated interaction between two large and complex magnetic environments only occurs between the two gas giants. This was first observed by Voyager 2 $\sim$ 40 years ago during its flyby of Saturn (e.g \citet{Desch1983, Kurth1982, Lepping1982, Lepping1983}) but no direct auroral measurements were taken. The idea behind our observations was to find whether the shocks produced from Saturn moving from rarefied plasma in Jupiter's magnetotail to the denser solar wind environment would be enough to dramatically compress Saturn's magnetosphere beyond the typical parameter space when Saturn is transitioning from one solar wind regime to another. Within Saturn's magnetosphere, the planetary plasma is dominated by heavy ions in (mainly \ion{H$_{2}$O}{$^{+}$}, \ion{OH}{$^{+}$}, \ion{O}{$^{+}$} produced from Saturn's icy moon Enceladus) \citep{Hui2010a}. This campaign afforded us to the opportunity to explore the extremes of parameter space at Saturn, and crucially the change between the typical water-rich kronian magnetosphere immersed in a variable solar wind, to a magnetosphere immersed in the rarefied jovian magnetotail. With the non-detection of Saturn throughout each of the observations, our analysis suggests that even with such a variable external driver, the dramatic compressions are still not enough to energise the heavy ions and produce X-ray aurorae. As stated by \citet{Hui2010a}, another likely possibility is that the field potentials at Saturn are too low to sufficiently charge strip magnetospheric plasma or to increase solar wind ion fluxes sufficiently to generate observable X-ray ion aurora \citep{Cravens2003, Clark2020}.  

From inspection of the two \textit{HST} observations taken within 2 days before ObsID 24847, a spiral-like morphology in the UV auroral oval is observed on November 22 and a more localised brightening near the pole is found hours before the Chandra observation. Models of the spiral morphology suggest that it is created from precipitation of hot plasma during intervals of steady yet unbalanced rapid reconnection in Saturn's magnetotail and during dayside magnetic reconnection \citep{Cowley2004, Jackman2004}. This latter type of reconnection may be associated with a shock, compressing the magnetosphere (e.g. possible shocks from the solar wind \citep{Badman2016a}) and is more likely to vary with solar wind dynamic pressure. One possible cause of the localised polar auroral brightening is from a dawn storm, auroral emissions from compression induced magnetic reconnection in Saturn's magnetotail \citep{Cowley2005}. Unfortunately without a significant X-ray detection during this time, we cannot determine whether there was any unusual X-ray emissions near the pole associated with the UV polar emissions.

This is the first X-ray campaign of its kind to look at a planet's magnetospheric response during such extreme conditions. We hope this work sets the foundations for further exploration into this rare planetary alignment in $\sim$ 20 years time with a new fleet of X-ray telescopes with greater spectral resolution from the \textit{X-IFU} on \textit{Athena} and \textit{Lynx's} \textit{HDXI}. This will allow tighter observational constraints on the auroral luminosity in all types of environments and add new insight into Saturn’s magnetospheric response. This will aid us to solve the mystery behind Saturn's absent X-ray aurora.

\section*{Acknowledgements}

We greatly thank Patrick Slane and the Chandra Scheduling Team for awarding us the Director's Discretionary Time and helping us to set up these unique observations. DMW would like to thank Sophie Murray for discussions on flaring activity during the early rising phase of the solar cycle and retrieving the GOES data used here. DMW would also like to thank Seán McEntee for discussions on solar X-ray scattering from the upper atmosphere of the gas giants. DMW is supported by the Science and Technology Facilities Council (STFC) studentship ST/S505703/1 and long term attachment grant to work at DIAS. WRD was supported by a STFC research grant to University College London (UCL). CMJ's work is supported by the Science foundation Ireland Grant 18/FRL/6199. ADW is supported by STFC studentship ST/S50578X/1. WRD and GB-R acknowledge support from STFC consolidated grant ST/S000240/1 to University College London (UCL).
\section*{Data Availability}

The Saturn X-ray observations we present here are publicly available from the \textit{Chandra Data Archive} (\url{https://cda.harvard.edu/chaser/}). The GOES data is publicly available from the NGDC archives (\url{https://www.ngdc.noaa.gov/stp/satellite/goes-r.html}) and can be retrieved and analysed via SunPy (\url{https://docs.sunpy.org/en/stable/whatsnew/2.1.html#overview}).




\bibliographystyle{mnras}
\bibliography{kronian_bib} 

\begin{thebibliography}{}
\makeatletter
\relax
\def\mn@urlcharsother{\let\do\@makeother \do\$\do\&\do\#\do\^\do\_\do\%\do\~}
\def\mn@doi{\begingroup\mn@urlcharsother \@ifnextchar [ {\mn@doi@}
  {\mn@doi@[]}}
\def\mn@doi@[#1]#2{\def\@tempa{#1}\ifx\@tempa\@empty \href
  {http://dx.doi.org/#2} {doi:#2}\else \href {http://dx.doi.org/#2} {#1}\fi
  \endgroup}
\def\mn@eprint#1#2{\mn@eprint@#1:#2::\@nil}
\def\mn@eprint@arXiv#1{\href {http://arxiv.org/abs/#1} {{\tt arXiv:#1}}}
\def\mn@eprint@dblp#1{\href {http://dblp.uni-trier.de/rec/bibtex/#1.xml}
  {dblp:#1}}
\def\mn@eprint@#1:#2:#3:#4\@nil{\def\@tempa {#1}\def\@tempb {#2}\def\@tempc
  {#3}\ifx \@tempc \@empty \let \@tempc \@tempb \let \@tempb \@tempa \fi \ifx
  \@tempb \@empty \def\@tempb {arXiv}\fi \@ifundefined
  {mn@eprint@\@tempb}{\@tempb:\@tempc}{\expandafter \expandafter \csname
  mn@eprint@\@tempb\endcsname \expandafter{\@tempc}}}

\bibitem[\protect\citeauthoryear{Ata{\c{c}} \&
  {\"{O}}zg{\"{u}}{\c{c}}}{Ata{\c{c}} \&
  {\"{O}}zg{\"{u}}{\c{c}}}{2001}]{Atac2001}
Ata{\c{c}} T.,  {\"{O}}zg{\"{u}}{\c{c}} A.,  2001, \mn@doi [Solar Physics]
  {10.1023/A:1005218315298}, 198, 399

\bibitem[\protect\citeauthoryear{Badman et~al.,}{Badman
  et~al.}{2011}]{Badman2011}
Badman S.~V.,  et~al., 2011, \mn@doi [Icarus] {10.1016/j.icarus.2011.09.031},
  216, 367

\bibitem[\protect\citeauthoryear{Badman, Branduardi-Raymont, Galand, Hess,
  Krupp, Lamy, Melin  \& Tao}{Badman et~al.}{2015}]{Badman2015}
Badman S.~V.,  Branduardi-Raymont G.,  Galand M.,  Hess S. L.~G.,  Krupp N.,
  Lamy L.,  Melin H.,   Tao C.,  2015, \mn@doi [Space Science Reviews]
  {10.1007/s11214-014-0042-x}, 187, 99

\bibitem[\protect\citeauthoryear{Badman et~al.,}{Badman
  et~al.}{2016}]{Badman2016a}
Badman S.~V.,  et~al., 2016, \mn@doi [Icarus] {10.1016/j.icarus.2014.11.014},
  263, 83

\bibitem[\protect\citeauthoryear{Bagenal}{Bagenal}{1992}]{Bagenal1992a}
Bagenal F.,  1992, Annual Review of Earth and Planetary Sciences, 20, 289

\bibitem[\protect\citeauthoryear{{Barret} et~al.,}{{Barret}
  et~al.}{2016}]{Barret2016}
{Barret} D.,  et~al., 2016, in {den Herder} J.-W.~A.,  {Takahashi} T.,
  {Bautz} M.,  eds,  Society of Photo-Optical Instrumentation Engineers (SPIE)
  Conference Series Vol. 9905, Space Telescopes and Instrumentation 2016:
  Ultraviolet to Gamma Ray. p. 99052F (\mn@eprint {arXiv} {1608.08105}),
  \mn@doi{10.1117/12.2232432}

\bibitem[\protect\citeauthoryear{Bhardwaj \& Gladstone}{Bhardwaj \&
  Gladstone}{2000}]{Bhardwaj2000a}
Bhardwaj A.,  Gladstone G.~R.,  2000, \mn@doi [Reviews of Geophysics]
  {10.1029/1998RG000046}, 38, 295

\bibitem[\protect\citeauthoryear{Bhardwaj et~al.,}{Bhardwaj
  et~al.}{2005a}]{Bhardwaj2005}
Bhardwaj A.,  et~al., 2005a, \mn@doi [Geophysical Research Letters]
  {10.1029/2004GL021497}, 32, 1

\bibitem[\protect\citeauthoryear{Bhardwaj, Elsner, {Waite, Jr.}, Gladstone,
  Cravens  \& Ford}{Bhardwaj et~al.}{2005b}]{Bhardwaj2005c}
Bhardwaj A.,  Elsner R.~F.,  {Waite, Jr.} J.~H.,  Gladstone G.~R.,  Cravens
  T.~E.,   Ford P.~G.,  2005b, \mn@doi [The Astrophysical Journal]
  {10.1086/430521}, 624, L121

\bibitem[\protect\citeauthoryear{Bhardwaj, Elsner, {Waite, Jr.}, Gladstone,
  Cravens  \& Ford}{Bhardwaj et~al.}{2005c}]{Bhardwaj2005b}
Bhardwaj A.,  Elsner R.~F.,  {Waite, Jr.} J.~H.,  Gladstone G.~R.,  Cravens
  T.~E.,   Ford P.~G.,  2005c, \mn@doi [The Astrophysical Journal]
  {10.1086/431933}, 627, L73

\bibitem[\protect\citeauthoryear{Blanc et~al.,}{Blanc et~al.}{2015}]{Blanc2015}
Blanc M.,  et~al., 2015, \mn@doi [Space Science Reviews]
  {10.1007/s11214-015-0172-9}, 192, 237

\bibitem[\protect\citeauthoryear{Bolton et~al.,}{Bolton
  et~al.}{2015}]{Bolton2015}
Bolton S.~J.,  et~al., 2015, {Jupiter's Magnetosphere: Plasma Sources and
  Transport}, \mn@doi{10.1007/s11214-015-0184-5}

\bibitem[\protect\citeauthoryear{Branduardi-Raymont, Elsner, Gladstone, Ramsay,
  Rodriguez, Soria  \& Waite}{Branduardi-Raymont et~al.}{2004}]{Elsner2004}
Branduardi-Raymont G.,  Elsner R.~F.,  Gladstone G.~R.,  Ramsay G.,  Rodriguez
  P.,  Soria R.,   Waite J.~H.,  2004, \mn@doi [Astronomy] {10.1051/0004-6361},
  337, 331

\bibitem[\protect\citeauthoryear{Branduardi-Raymont et~al.,}{Branduardi-Raymont
  et~al.}{2007a}]{Branduardi-Raymont2007a}
Branduardi-Raymont G.,  et~al., 2007a, \mn@doi [Planetary and Space Science]
  {10.1016/j.pss.2006.11.017}, 55, 1126

\bibitem[\protect\citeauthoryear{Branduardi-Raymont et~al.,}{Branduardi-Raymont
  et~al.}{2007b}]{Branduardi-Raymont2007b}
Branduardi-Raymont G.,  et~al., 2007b, \mn@doi [Astronomy and Astrophysics]
  {10.1051/0004-6361:20066406}, 463, 761

\bibitem[\protect\citeauthoryear{Branduardi-Raymont, Elsner, Galand, Grodent,
  Cravens, Ford, Gladstone  \& Waite}{Branduardi-Raymont
  et~al.}{2008}]{Branduardi-Raymont2008}
Branduardi-Raymont G.,  Elsner R.~F.,  Galand M.,  Grodent D.,  Cravens T.~E.,
  Ford P.,  Gladstone G.~R.,   Waite J.~H.,  2008, \mn@doi [Journal of
  Geophysical Research: Space Physics] {10.1029/2007JA012600}, 113, 1

\bibitem[\protect\citeauthoryear{Branduardi-Raymont, Bhardwaj, Elsner  \&
  Rodriguez}{Branduardi-Raymont et~al.}{2010}]{Branduardi-Raymont2010}
Branduardi-Raymont G.,  Bhardwaj A.,  Elsner R.~F.,   Rodriguez P.,  2010,
  \mn@doi [Astronomy and Astrophysics] {10.1051/0004-6361/200913110}, 510, 1

\bibitem[\protect\citeauthoryear{Branduardi-Raymont et~al.,}{Branduardi-Raymont
  et~al.}{2013}]{Branduardi-Raymont2013}
Branduardi-Raymont G.,  et~al., 2013, \mn@doi [Journal of Geophysical Research:
  Space Physics] {10.1002/jgra.50112}, 118, 2145

\bibitem[\protect\citeauthoryear{Broadfoot et~al.,}{Broadfoot
  et~al.}{1981}]{Broadfoot1981}
Broadfoot A.~L.,  et~al., 1981, \mn@doi [Science]
  {10.1126/science.212.4491.206}, 212, 206

\bibitem[\protect\citeauthoryear{Burlaga, Skoug, Smith, Webb, Zurbuchen  \&
  Reinard}{Burlaga et~al.}{2001}]{Burlaga2001}
Burlaga L.~F.,  Skoug R.~M.,  Smith C.~W.,  Webb D.~F.,  Zurbuchen T.~H.,
  Reinard A.,  2001, \mn@doi [Journal of Geophysical Research: Space Physics]
  {10.1029/2000ja000214}, 106, 20957

\bibitem[\protect\citeauthoryear{Carbary \& Krimigis}{Carbary \&
  Krimigis}{1983}]{Carbary1983}
Carbary J.~F.,  Krimigis S.~M.,  1983, \mn@doi [Journal of Geophysical
  Research] {10.1029/JA088iA11p08947}, 88, 8947

\bibitem[\protect\citeauthoryear{Clark et~al.,}{Clark et~al.}{2020}]{Clark2020}
Clark G.,  et~al., 2020, \mn@doi [Journal of Geophysical Research: Space
  Physics] {10.1029/2020ja028052}, 125, 1

\bibitem[\protect\citeauthoryear{Cowley, Bunce  \& O'Rourke}{Cowley
  et~al.}{2004}]{Cowley2004}
Cowley S. W.~H.,  Bunce E.~J.,   O'Rourke J.~M.,  2004, \mn@doi [Journal of
  Geophysical Research: Space Physics] {10.1029/2003JA010375}, 109

\bibitem[\protect\citeauthoryear{Cowley et~al.,}{Cowley
  et~al.}{2005}]{Cowley2005}
Cowley S.~W.,  et~al., 2005, \mn@doi [Journal of Geophysical Research: Space
  Physics] {10.1029/2004JA010796}, 110, 1

\bibitem[\protect\citeauthoryear{Cravens, Waite, Gombosi, Lugaz, Gladstone,
  Mauk  \& MacDowall}{Cravens et~al.}{2003}]{Cravens2003}
Cravens T.~E.,  Waite J.~H.,  Gombosi T.~I.,  Lugaz N.,  Gladstone G.~R.,  Mauk
  B.~H.,   MacDowall R.~J.,  2003, \mn@doi [Journal of Geophysical Research:
  Space Physics] {10.1029/2003JA010050}, 108, 1

\bibitem[\protect\citeauthoryear{Desch}{Desch}{1983}]{Desch1983}
Desch M.~D.,  1983, \mn@doi [Journal of Geophysical Research]
  {10.1029/JA088iA09p06904}, 88, 6904

\bibitem[\protect\citeauthoryear{Dunn et~al.,}{Dunn et~al.}{2016}]{Dunn2016a}
Dunn W.~R.,  et~al., 2016, \mn@doi [Journal of Geophysical Research A: Space
  Physics] {10.1002/2015JA021888}, 121, 2274

\bibitem[\protect\citeauthoryear{Dunn et~al.,}{Dunn et~al.}{2017}]{Dunn2017}
Dunn W.~R.,  et~al., 2017, \mn@doi [Nature Astronomy]
  {10.1038/s41550-017-0262-6}, 1, 758

\bibitem[\protect\citeauthoryear{Dunn et~al.,}{Dunn et~al.}{2020}]{Dunn2020a}
Dunn W.~R.,  et~al., 2020, \mn@doi [Journal of Geophysical Research: Space
  Physics] {10.1029/2019JA027219}, 125, e2019JA027219

\bibitem[\protect\citeauthoryear{Dunn et~al.,}{Dunn et~al.}{2021}]{Dunn2021}
Dunn W.~R.,  et~al., 2021, \mn@doi [Journal of Geophysical Research: Space
  Physics] {10.1029/2020JA028739}, 126

\bibitem[\protect\citeauthoryear{Elsner et~al.,}{Elsner
  et~al.}{2005}]{Elsner2005}
Elsner R.~F.,  et~al., 2005, \mn@doi [Journal of Geophysical Research: Space
  Physics] {10.1029/2004JA010717}, 110, 1

\bibitem[\protect\citeauthoryear{Falcone, Kraft, Bautz, Gaskin, Mulqueen  \&
  Swartz}{Falcone et~al.}{2019}]{Falcone2019}
Falcone A.~D.,  Kraft R.~P.,  Bautz M.~W.,  Gaskin J.~A.,  Mulqueen J.~A.,
  Swartz D.~A.,  2019, \mn@doi [Journal of Astronomical Telescopes,
  Instruments, and Systems] {10.1117/1.JATIS.5.2.021019}, 5

\bibitem[\protect\citeauthoryear{Fletcher et~al.,}{Fletcher
  et~al.}{2011}]{Fletcher2011}
Fletcher L.,  et~al., 2011, \mn@doi [Space Science Reviews]
  {10.1007/s11214-010-9701-8}, 159, 19

\bibitem[\protect\citeauthoryear{Gaskin et~al.,}{Gaskin
  et~al.}{2019}]{Gaskin2019}
Gaskin J.~A.,  et~al., 2019, \mn@doi [Journal of Astronomical Telescopes,
  Instruments, and Systems] {10.1117/1.JATIS.5.2.021001}, 5, 1

\bibitem[\protect\citeauthoryear{Gilman, Hurley, Seward, Schnopper, Sullivan
  \& Metzger}{Gilman et~al.}{1986}]{Gilman1986}
Gilman D.~A.,  Hurley K.~C.,  Seward F.~D.,  Schnopper H.~W.,  Sullivan J.~D.,
   Metzger A.~E.,  1986, \mn@doi [The Astrophysical Journal] {10.1086/163820},
  300, 453

\bibitem[\protect\citeauthoryear{Gladstone et~al.,}{Gladstone
  et~al.}{2002}]{Gladstone2002}
Gladstone G.~R.,  et~al., 2002, \mn@doi [Nature] {10.1038/4151000a}, 415, 1000

\bibitem[\protect\citeauthoryear{Goodman et~al.,}{Goodman
  et~al.}{2013}]{Goodman2013}
Goodman S.~J.,  et~al., 2013, \mn@doi [Atmospheric Research]
  {10.1016/j.atmosres.2013.01.006}, 125-126, 34

\bibitem[\protect\citeauthoryear{Gurnett, KURTH  \& SCARF}{Gurnett
  et~al.}{1979}]{Gurnett1979a}
Gurnett D.~A.,  KURTH W.~S.,   SCARF F.~L.,  1979, \mn@doi [Science]
  {10.1126/science.206.4421.987}, 206, 987

\bibitem[\protect\citeauthoryear{Hui, Cravens, Ozak  \& Schultz}{Hui
  et~al.}{2010}]{Hui2010a}
Hui Y.,  Cravens T.~E.,  Ozak N.,   Schultz D.~R.,  2010, \mn@doi [Journal of
  Geophysical Research: Space Physics] {10.1029/2010JA015639}, 115

\bibitem[\protect\citeauthoryear{Jackman \& Arridge}{Jackman \&
  Arridge}{2011}]{Jackman2011}
Jackman C.~M.,  Arridge C.~S.,  2011, \mn@doi [Solar Physics]
  {10.1007/s11207-011-9748-z}, 274, 481

\bibitem[\protect\citeauthoryear{Jackman, Achilleos, Bunce, Cowley, Dougherty,
  Jones, Milan  \& Smith}{Jackman et~al.}{2004}]{Jackman2004}
Jackman C.~M.,  Achilleos N.,  Bunce E.~J.,  Cowley S.~W.,  Dougherty M.~K.,
  Jones G.~H.,  Milan S.~E.,   Smith E.~J.,  2004, \mn@doi [Journal of
  Geophysical Research: Space Physics] {10.1029/2004JA010614}, 109

\bibitem[\protect\citeauthoryear{Jackman et~al.,}{Jackman
  et~al.}{2018}]{Jackman2018a}
Jackman C.~M.,  et~al., 2018, \mn@doi [Journal of Geophysical Research: Space
  Physics] {10.1029/2018JA025490}, 123, 9204

\bibitem[\protect\citeauthoryear{Joye \& Mandel}{Joye \&
  Mandel}{2003}]{Joye2003}
Joye W.,  Mandel E.,  2003, Astronomical Data Analysis Software and Systems
  XII, 295, 489

\bibitem[\protect\citeauthoryear{Kurth, Sullivan, Gurnett, Scarf, Bridge  \&
  Sittler}{Kurth et~al.}{1982}]{Kurth1982}
Kurth W.~S.,  Sullivan J.~D.,  Gurnett D.~A.,  Scarf F.~L.,  Bridge H.~S.,
  Sittler E.~C.,  1982, \mn@doi [Journal of Geophysical Research]
  {10.1029/JA087iA12p10373}, 87, 10373

\bibitem[\protect\citeauthoryear{Kurth et~al.,}{Kurth et~al.}{2005}]{Kurth2005}
Kurth W.~S.,  et~al., 2005, \mn@doi [Nature] {10.1038/nature03334}, 433, 722

\bibitem[\protect\citeauthoryear{Lamy, Cecconi, Prang{\'{e}}, Zarka, Nichols
  \& Clarke}{Lamy et~al.}{2009}]{Lamy2009}
Lamy L.,  Cecconi B.,  Prang{\'{e}} R.,  Zarka P.,  Nichols J.~D.,   Clarke
  J.~T.,  2009, \mn@doi [Journal of Geophysical Research: Space Physics]
  {10.1029/2009JA014401}, 114, 1

\bibitem[\protect\citeauthoryear{Lemen et~al.,}{Lemen et~al.}{2004}]{Lemen2004}
Lemen J.~R.,  et~al., 2004, \mn@doi [Proceedings of SPIE Vol. 5171 (Telescopes
  and Instrumentation for Solar Astrophysics)] {10.1117/12.507566}, 5171, 65

\bibitem[\protect\citeauthoryear{Lepping, Burlaga, Desch  \& Klein}{Lepping
  et~al.}{1982}]{Lepping1982}
Lepping R.~P.,  Burlaga L.~F.,  Desch M.~D.,   Klein L.~W.,  1982, \mn@doi
  [Geophysical Research Letters] {10.1029/GL009i008p00885}, 9, 885

\bibitem[\protect\citeauthoryear{Lepping, Desch, Klein, Sittler, Sullivan,
  Kurth  \& Behannon}{Lepping et~al.}{1983}]{Lepping1983}
Lepping R.~P.,  Desch M.~D.,  Klein L.~W.,  Sittler E.~C.,  Sullivan J.~D.,
  Kurth W.~S.,   Behannon K.~W.,  1983, \mn@doi [Journal of Geophysical
  Research: Space Physics] {10.1029/JA088iA11p08801}, 88, 8801

\bibitem[\protect\citeauthoryear{McComas, Allegrini, Bagenal, Crary, Ebert,
  Elliott, Stern  \& Valek}{McComas et~al.}{2007}]{Mccomas2007}
McComas D.~J.,  Allegrini F.,  Bagenal F.,  Crary F.,  Ebert R.~W.,  Elliott
  H.,  Stern A.,   Valek P.,  2007, \mn@doi [Science]
  {10.1126/science.1147393}, 318, 217

\bibitem[\protect\citeauthoryear{Metzger, Gilman, Luthey, Hurley, Schnopper,
  Seward  \& Sullivan}{Metzger et~al.}{1983}]{Metzger1983}
Metzger A.~E.,  Gilman D.~A.,  Luthey J.~L.,  Hurley K.~C.,  Schnopper H.~W.,
  Seward F.~D.,   Sullivan J.~D.,  1983, \mn@doi [Journal of Geophysical
  Research] {10.1029/JA088iA10p07731}, 88, 7731

\bibitem[\protect\citeauthoryear{Ness \& Schmitt}{Ness \&
  Schmitt}{2000}]{Ness2000}
Ness J.~U.,  Schmitt J.~H.,  2000, Astronomy and Astrophysics, 355, 394

\bibitem[\protect\citeauthoryear{Ness, Schmitt  \& Robrade}{Ness
  et~al.}{2004a}]{Ness2004a}
Ness J.~U.,  Schmitt J.~H.,   Robrade J.,  2004a, \mn@doi [Astronomy and
  Astrophysics] {10.1051/0004-6361:20031761}, 414, 49

\bibitem[\protect\citeauthoryear{Ness, Schmitt, Wolk, Dennerl  \& Burwitz}{Ness
  et~al.}{2004b}]{Ness2004}
Ness J.~U.,  Schmitt J.~H.,  Wolk S.~J.,  Dennerl K.,   Burwitz V.,  2004b,
  \mn@doi [Astronomy and Astrophysics] {10.1051/0004-6361:20035736}, 418, 337

\bibitem[\protect\citeauthoryear{Nichols et~al.,}{Nichols
  et~al.}{2010}]{Nichols2010}
Nichols J.~D.,  et~al., 2010, \mn@doi [Geophysical Research Letters]
  {10.1029/2010GL044057}, 37, 1

\bibitem[\protect\citeauthoryear{Nichols, Badman, Bunce, Clarke, Cowley, Hunt
  \& Provan}{Nichols et~al.}{2016}]{Nichols2016}
Nichols J.~D.,  Badman S.~V.,  Bunce E.~J.,  Clarke J.~T.,  Cowley S.~W.,  Hunt
  G.~J.,   Provan G.,  2016, \mn@doi [Icarus] {10.1016/j.icarus.2015.09.008},
  263, 17

\bibitem[\protect\citeauthoryear{Opp}{Opp}{1980}]{Opp1980}
Opp A.~G.,  1980, \mn@doi [Science] {10.1126/science.207.4429.401}, 207, 401

\bibitem[\protect\citeauthoryear{Ozak, Schultz, Cravens, Kharchenko  \&
  Hui}{Ozak et~al.}{2010}]{Ozak2010}
Ozak N.,  Schultz D.~R.,  Cravens T.~E.,  Kharchenko V.,   Hui Y.~W.,  2010,
  \mn@doi [Journal of Geophysical Research: Space Physics]
  {10.1029/2010JA015635}, 115, 1

\bibitem[\protect\citeauthoryear{{Plucinsky}, {Bogdan}, {Marshall}  \&
  {Tice}}{{Plucinsky} et~al.}{2018}]{Plucinsky2018}
{Plucinsky} P.~P.,  {Bogdan} A.,  {Marshall} H.~L.,   {Tice} N.~W.,  2018, in
  {den Herder} J.-W.~A.,  {Nikzad} S.,   {Nakazawa} K.,  eds,  Society of
  Photo-Optical Instrumentation Engineers (SPIE) Conference Series Vol. 10699,
  Space Telescopes and Instrumentation 2018: Ultraviolet to Gamma Ray. p.
  106996B (\mn@eprint {arXiv} {1809.02225}), \mn@doi{10.1117/12.2312748}

\bibitem[\protect\citeauthoryear{Reed, Jackman, Lamy, Kurth  \& Whiter}{Reed
  et~al.}{2018}]{Reed2018}
Reed J.~J.,  Jackman C.~M.,  Lamy L.,  Kurth W.~S.,   Whiter D.~K.,  2018,
  \mn@doi [Journal of Geophysical Research: Space Physics]
  {10.1002/2017JA024499}, 123, 443

\bibitem[\protect\citeauthoryear{Sandel et~al.,}{Sandel
  et~al.}{1982}]{SANDEL1982}
Sandel B.~R.,  et~al., 1982, \mn@doi [Science] {10.1126/science.215.4532.548},
  215, 548

\bibitem[\protect\citeauthoryear{Stallard, Miller, Melin, Lystrup, Cowley,
  Bunce, Achilleos  \& Dougherty}{Stallard et~al.}{2008a}]{Stallard2008}
Stallard T.,  Miller S.,  Melin H.,  Lystrup M.,  Cowley S. W.~H.,  Bunce
  E.~J.,  Achilleos N.,   Dougherty M.,  2008a, \mn@doi [Nature]
  {10.1038/nature07077}, 453, 1083

\bibitem[\protect\citeauthoryear{Stallard et~al.,}{Stallard
  et~al.}{2008b}]{Stallard2008a}
Stallard T.,  et~al., 2008b, \mn@doi [Nature] {10.1038/nature07440}, 456, 214

\bibitem[\protect\citeauthoryear{Weigt et~al.,}{Weigt et~al.}{2020}]{Weigt2020}
Weigt D.~M.,  et~al., 2020, \mn@doi [Journal of Geophysical Research: Planets]
  {10.1029/2019JE006262}, 125, e2019JE006262

\bibitem[\protect\citeauthoryear{Weigt, Jackman, Vogt, Manners, Dunn,
  Gladstone, Kraft  \& Branduardi-raymont}{Weigt et~al.}{2021}]{Weigt2021}
Weigt D.~M.,  Jackman C.~M.,  Vogt M.~F.,  Manners H.,  Dunn W.~R.,  Gladstone
  G.~R.,  Kraft R.,   Branduardi-raymont G.,  2021, \mn@doi [Earth and Space
  Science Open Archive] {10.1002/essoar.10506198.1}, p.~40

\bibitem[\protect\citeauthoryear{Wibisono et~al.,}{Wibisono
  et~al.}{2020}]{Wibisono2020}
Wibisono A.~D.,  et~al., 2020, \mn@doi [Journal of Geophysical Research: Space
  Physics] {10.1029/2019JA027676}, 125, e2019JA027676

\bibitem[\protect\citeauthoryear{Xystouris, Sigala  \&
  Mavromichalaki}{Xystouris et~al.}{2014}]{Xystouris2014}
Xystouris G.,  Sigala E.,   Mavromichalaki H.,  2014, \mn@doi [Solar Physics]
  {10.1007/s11207-013-0355-z}, 289, 995

\bibitem[\protect\citeauthoryear{Zarka}{Zarka}{1998}]{Zarka1998}
Zarka P.,  1998, \mn@doi [Journal of Geophysical Research: Planets]
  {10.1029/98JE01323}, 103, 20159

\makeatother
\end{thebibliography}








\bsp	
\label{lastpage}
\end{document}